\documentclass[sigconf]{acmart}
\usepackage{multirow}
\usepackage{colortbl} 
\usepackage{xcolor}   
\usepackage{enumitem}
\usepackage{subfigure} 

\copyrightyear{2026}
\acmYear{2026}
\setcopyright{cc}
\setcctype{by}
\acmConference[SIGIR '26]{Proceedings of the 49th International ACM SIGIR Conference on Research and Development in Information Retrieval}{July 20--24, 2026}{Melbourne, VIC, Australia}
\acmBooktitle{Proceedings of the 49th International ACM SIGIR Conference on Research and Development in Information Retrieval (SIGIR '26), July 20--24, 2026, Melbourne, VIC, Australia}
\acmDOI{10.1145/3805712.3809552}
\acmISBN{979-8-4007-2599-9/2026/07}

\begin{document}
\title{Discrete Preference Learning for Personalized Multimodal Generation}

\author{Yuting Zhang}
\email{yzhang755@connect.hkust-gz.edu.cn}
\affiliation{
  \department{Thrust of Artificial Intelligence}
  \institution{Hong Kong University of Science and Technology (Guangzhou)}
  \city{Guangzhou}
  \state{Guangdong}
  \country{China}
}

\author{Ying Sun}
\authornote{Corresponding authors.}
\email{yings@hkust-gz.edu.cn}
\affiliation{
  \department{Thrust of Artificial Intelligence}
  \institution{Hong Kong University of Science and Technology (Guangzhou)}
  \city{Guangzhou}
  \state{Guangdong}
  \country{China}
}

\author{Dazhong Shen}
\email{shendazhong@nuaa.edu.cn}
\affiliation{%
  \institution{Nanjing University of Aeronautics and Astronautics}
 \city{Nanjing}
  \state{Jiangsu}
  \country{China}
}

\author{Ziwei Xie}
\email{xieziwei@oppo.com}
\affiliation{%
  \institution{OPPO Research Institute}
  \city{Shenzhen}
  \state{Guangdong}
  \country{China}
}

\author{Feng Liu}
\email{liufeng4hit@gmail.com}
\affiliation{%
  \institution{OPPO Research Institute}
  \city{Shenzhen}
  \state{Guangdong}
  \country{China}
}

\author{Changwang Zhang}
\authornotemark[1]
\email{changwangzhang@foxmail.com}
\affiliation{%
  \institution{OPPO Research Institute}
  \city{Shenzhen}
  \state{Guangdong}
  \country{China}
}

\author{Xiang Liu}
\email{liuxiang10@oppo.com}
\affiliation{%
  \institution{OPPO Internet Services System}
  \city{Shenzhen}
  \state{Guangdong}
  \country{China}
}

\author{Jun Wang}
\authornotemark[1]
\email{junwang.lu@gmail.com}
\affiliation{%
  \institution{OPPO Research Institute}
  \city{Shenzhen}
  \state{Guangdong}
  \country{China}
}

\author{Hui Xiong}
\authornotemark[1]
\email{xionghui@hkust-gz.edu.cn}
\affiliation{%
  \department{Thrust of Artificial Intelligence}
  \institution{Hong Kong University of Science and Technology (Guangzhou)}
  \city{Guangzhou}
  \state{Guangdong}
  \country{China}
}

\renewcommand{\shortauthors}{Yuting Zhang et al.}

\begin{abstract}

The emergence of generative models enables the creation of texts and images tailored to users' preferences.  Existing personalized generative models have two critical limitations:   lacking a dedicated paradigm for accurate preference modeling, and generating unimodal content despite real-world multimodal-driven user interactions. Therefore, we propose personalized multimodal generation, which captures modal-specific preferences via a dedicated preference model from multimodal interactions, and then feeds them into downstream generators for personalized multimodal content. However, this task presents two challenges:  (1) Gap between \textbf{continuous} preferences from dedicated modeling and \textbf{discrete} token inputs intrinsic to generator architectures; (2) Potential \textbf{inconsistency} between generated images and texts. To tackle these, we present a two-stage framework called \underline{D}iscrete \underline{P}reference learning for \underline{P}ersonalized \underline{M}ultimodal \underline{G}eneration (DPPMG).
In the first stage, to accurately learn discrete modal-specific preferences, we introduce a modal-specific graph neural network (a dedicated preference model) to learn users' modal-specific preferences, which preferences are then quantized into discrete preference tokens. In the second stage,  the discrete modal-specific preference tokens are injected into downstream text and image generators.  To further enhance cross-modal consistency while preserving personalization, we design a cross-modal consistent and personalized reward to fine-tune token-associated parameters. Extensive experiments on two real-world datasets demonstrate the effectiveness of our model in generating personalized and consistent multimodal content.

\end{abstract}



\maketitle
\section{Introduction}
Aiming to infer user preferences from historical behaviors and provide tailored content accordingly, personalized systems (e.g., recommenders~\cite{zhang2019deep}) have played a crucial role in various online platforms. However, most personalized systems focus on mining user preferences and retrieving existing contents accordingly~\cite{wu2024survey}.
The rapid development of generative models, such as LLMs~\cite{achiam2023gpt,grattafiori2024llama} and Diffusion models (DMs) ~\cite{rombach2022high,podellsdxl}, has spurred the emergence of a personalized generation paradigm. It directly generates content tailored to individual preferences, which is becoming increasingly important across various domains, including movie marketing~\cite{shen2024pmg,xu2025personalized}, advertising~\cite{yang2024new,chen2025ctr}, music~\cite{dai2022personalised, plitsis2024investigating}, and fashion designs~\cite{xu2024diffusion}.

Nevertheless, existing personalized generation methods struggle to accurately extract user preferences. For instance, personalized text generation~\cite{salemi2024optimization,tan2024personalized} typically relies on retrieving text from users' interaction history for generation guidance. The LaMP benchmark~\cite{salemi2023lamp} evaluates several retrieval-augmented strategies, including BM25~\cite{robertson1995okapi}, Recency and Contriever~\cite{izacard2021unsupervised}. However, retrieved text often contains noise, mixing users' preferred and dispreferred features (e.g., style, word choice), hindering preference-aligned text generation.
The same issue also persists in personalized image generation. Existing works directly use generation models (DMs~\cite{galimage,ruiz2023dreambooth,xu2024diffusion}, LLMs~\cite{shen2024pmg} or MLLMs~\cite{xu2025personalized}) to learn preferences from users' historical images. For instance, PMG~\cite{shen2024pmg} converts historical images into textual descriptions and employs generative LLMs to extract keywords for guiding image generation.  However, these \textbf{generative approaches lack a dedicated preference learning paradigm}, such as classical collaborative filtering that mines user-item interactions and user/item similarity patterns~\cite{su2009survey}. Consequently, they fail to accurately extract user preferences from complex historical data.

Furthermore, users’ content demands are inherently multimodal, as evidenced by the multimodal content deployed across mainstream platforms to capture attention. For example, advertising platforms provide visual images with textual copy; movie marketing pairs posters with introductions; short videos integrate text, images, and audio elements. These modalities collectively drive user engagement~\cite{liu2024multimodal,xu2025survey}, making \textbf{the isolated single-modal generation paradigms inherently inadequate}.

Thus, we introduce the personalized multimodal generation task, which technically aims to capture users’ modality-specific preferences from multimodal interaction histories and leverage these to generate tailored multimodal content (e.g., text-image pairs). However, this task poses two major challenges:

(1) How to accurately extract user preferences and effectively incorporate them into downstream generators. Traditional preference models such as graph neural networks~\cite{he2020lightgcn, wu2022graph, he2017neural} excel at extracting users'  preferences using collaborative signals but represent these preferences as continuous embeddings. In contrast, current mainstream generative models (e.g., LLMs with \textbf{Transformer} backbones and diffusion models with \textbf{attention} layers) inherently rely on discrete sequences as input, creating a fundamental gap.

\begin{figure}
    \centering
    \includegraphics[width=0.9\linewidth]{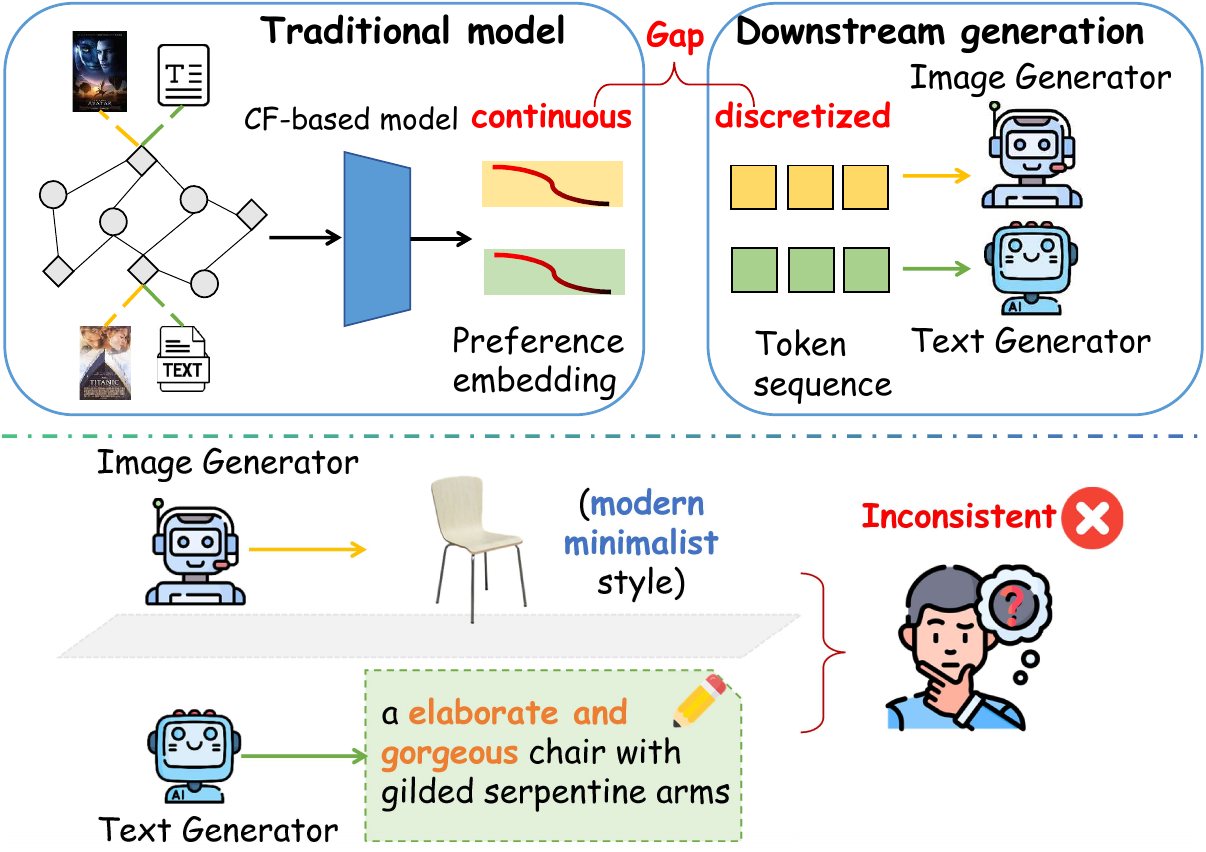}
    \caption{Challenges faced by personalized multimodal content generation: (1) Gap between continuous preference representations and generators' discrete token sequence; (2) Inconsistency between generated images and texts.}
    \label{fig:challenges}
\end{figure}

(2) How to realize consistent personalized multimodal generation.  Performing single-modal generation~\cite{salemi2023lamp,salemi2024optimization,tan2024personalized} separately might \textbf{neglect cross-modal alignment}, resulting in disjointed image-text semantics and inconsistent styles. For example, as Figure~\ref{fig:challenges} shows, a generated image depicting a piece of modern minimalist furniture may be paired with an elaborate, gorgeous textual description, which will directly degrade user experience and satisfaction.

To address these challenges, we propose a Discrete Preference Learning framework for Personalized Multimodal Generation. The framework comprises two core stages:
(1) We first introduce a modal-aware graph neural network to accurately model users’ modal-specific preference representations, and then quantize them into discrete preference tokens. This process is guided by specially designed cross-modal and preference-oriented tasks.
(2) We then integrate the discretized modal-specific preference tokens into the downstream text and image generators. To further enhance cross-modal consistency while preserving personalization, we design a cross-modal and personalized reward function to fine-tune preference token-related parameters.
To summarize, our main contributions are listed below:
\begin{itemize}[leftmargin=*,itemsep=2pt,topsep=0pt,parsep=0pt]
    \item To our knowledge, we are among the first to explicitly enable personalized multimodal generation, which captures users' modality-aware preferences from multimodal interaction history data.
    \item We propose discrete preference token learning, which adopts vector quantization to align continuous preference embeddings with discrete inputs natively required by generative architectures.
    \item During training, we freeze the pre-trained generative backbones and optimize lightweight token-related parameters as additional generation conditions, ensuring high training efficiency.
    \item Extensive experiments and human evaluation on two real-world datasets verify the effectiveness of our model in generating personalized, consistent multimodal content.
\end{itemize}

\section{Related Work}
\subsection{Multimodal Content Generation}
Multimodal content generation targets cross-modal  consistent and preference-aligned content creation.  Existing works~\cite{achiam2023gpt,team2023gemini,hou2025personavlog} feed \textbf{explicit} instructions (e.g., Pixar Style) for instruction-aligned generation. In contrast,  our personalized generation infers \textbf{implicit} preferences from users’ multimodal historical interactions, analogous to recommenders \cite{zhang2019deep,he2020lightgcn,sun2025market}, and generates consistent preference-aligned \textbf{novel} content.  Few studies explicitly generate personalized multimodal content while ensuring consistency. We thus review unimodal personalized generation (text and image) respectively:

\textbf{Personalized Text Generation}. It has been widely applied in various applications, such as dialogue agents~\cite{zhong2022less,vincent2023personalised}, headline~\cite{ao2021pens} and product description generation~\cite{elad2019learning,chen2019towards}. With the rise of LLMs~\cite{grattafiori2024llama,achiam2023gpt} applied in various scenarios~\cite{shen2024pmg,cui2026llm,
qin2026large,zhang2026enhancing,xin2025llmcdsr}, benchmarks such as LaMP~\cite{salemi2023lamp} evaluate different retrieval-augmented strategies (e.g., BM25~\cite{robertson1995okapi}, Recency, Contriever~\cite{izacard2021unsupervised}) for personalized text generation. Further, RSPG~\cite{salemi2024optimization} uses meta-learning to select optimal retrieval methods, while PER-PCS~\cite{tan2024personalized} introduces a parameter-sharing framework for efficient personalization.

\textbf{Personalized Image Generation}. Existing works primarily rely on DMs, LLMs or increasingly MLLMs. (1) DM-based methods such as Textual Inversion~\cite{galimage} and DreamBooth~\cite{ruiz2023dreambooth} align generation with explicit user instructions but rarely capture implicit visual preferences. DiFashion~\cite{xu2024diffusion}, CG4CTR~\cite{yang2024new}, and AdBooster~\cite{shilova2023adbooster} further fine-tune diffusion models on user-interaction data to personalize fashion and product images.
(2) LLM/MLLM-based methods such as PMG~\cite{shen2024pmg} convert images into text to infer preferences, but inevitably lose rich visual details. Pigeon~\cite{xu2025personalized} instead exploits MLLMs to extract preferences directly from users’ historical images. 

Separate personalized text-image generation risks cross-modal inconsistency (e.g., luxury captions mismatched with minimalist images).
To address this, we propose a cross-modal alignment reward for coherent generation. Additionally, instead of extending prior methods, we introduce preference-oriented vector quantization to encode preferences as discrete tokens for downstream generation.

\subsection{Vector Quantization for Generation}
Vector Quantization (VQ)~\cite{linde2003algorithm} has advanced significantly in generative modeling, starting with VQ-VAE \cite{van2017neural}, which maps continuous latents to discrete codebook entries for efficient compression and generation. Product Quantization \cite{jegou2010product,chen2020differentiable,el2022image} improves scalability by decomposing the feature space into subspaces, reducing complexity. Residual Quantization (RQ) \cite{martinez2014stacked,zeghidour2021soundstream,lee2022autoregressive} further enhances fidelity via iterative residual encoding, enabling hierarchical representations. More recently, Lookup-free VQ \cite{mentzerfinite,yu2023language} eliminates codebook lookups by quantizing each dimension independently, boosting efficiency in sequence modeling. Despite these advances, most VQ methods focus on reconstruction but lack the integration of preference information, limiting personalization in generation.

\section{Preliminary}
We aim to extract users’ visual and textual preferences from their interaction histories and feed them into downstream generators to create personalized multimodal content. We thus first collect users’ multimodal interaction data, defined as follows:

Let $\mathcal{U}$ and $\mathcal{I}$ denote the universal user and item sets. Users’ interactions (e.g., viewing, clicking) with items form a multimodal bipartite graph $\mathcal{G}=\{(u,i)|u\in \mathcal{U},i\in \mathcal{I}\}$, where each item node has visual and textual features. To capture modal-specific user preferences, we process each modality independently instead of unifying multimodal information, and construct modality-specific bipartite interaction graphs as follows:

\textbf{\textit{Modal-Specific Bipartite Interaction  Graphs}}. For simplicity, we adopt $m \in \{v, t\}$ as the modality indicator, where $v,  t$ represent the visual and textual modalities, respectively. We split $\mathcal{G}$ into two modality-specific bipartite graphs $\mathcal{G}^m=\{(u,i^m) | (u,i) \in \mathcal{G}\}$, where $i^m$ retains only modality-$m$ features of the original item $i$.

Given the above definitions, multimodal personalized content generation is defined as follows:

\textbf{\textit{Multimodal Personalized Content Generation}}. Given modal-specific bipartite interaction graphs $\mathcal{G}^m$, a target user $u$ and a reference item $i$, our goal is to generate personalized multimodal content, including image and text modalities for $i$, that not only aligns with $u$'s visual and textual preferences but also remains consistent with $i$'s original multimodal content.

This task has significant practical value for real-world applications, directly enhancing user experience and engagement.
\begin{itemize}[leftmargin=*,itemsep=2pt,topsep=0pt,parsep=0pt]
\item Advertising: Automatically generates personalized advertising images and copy that align with user preferences while retaining core product messaging, boosting click-through rates.
\item E-commerce: Produces tailored product images and descriptions matching user styles, helping reduce decision-making friction.
\item Media entertainment: Creates movie posters and introductions aligned with user preferences, increasing user stickiness.
\end{itemize}

\section{Methodology}

\begin{figure*}
    \centering
    \includegraphics[width=0.7\linewidth]{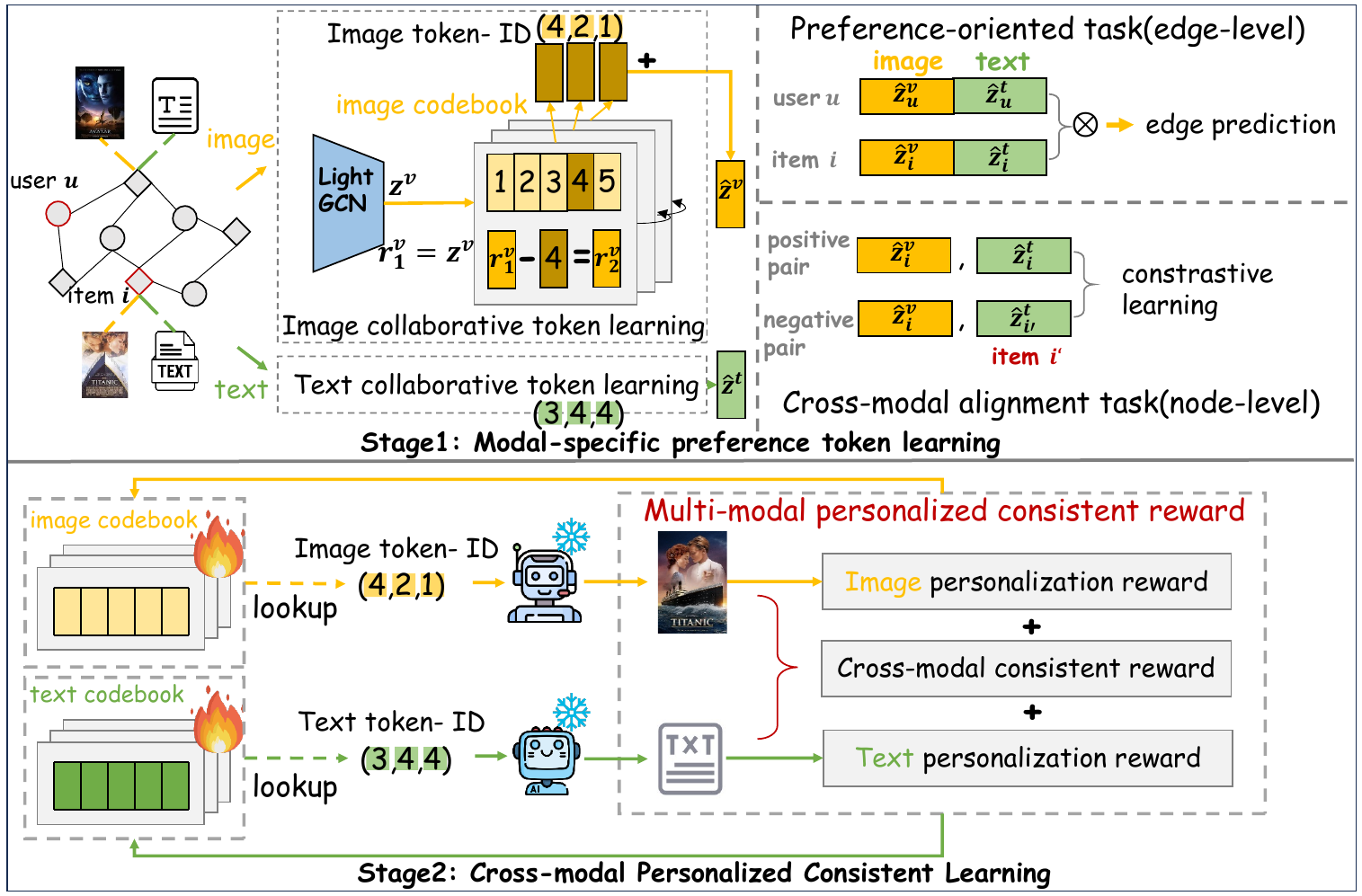}
    \caption{Overview of DPPMG. Stage 1 learns modal-specific preference tokens via edge-level preference-oriented and node-level cross-modal alignment tasks. Stage 2 injects tokens into downstream models, optimized by a cross-modal personalized consistency reward. The text collaborative token learning follows the same logic as the image branch. For brevity, only token IDs are shown, with generator instructions omitted.}
    \label{fig:framework}
\end{figure*}

This section details DPPMG, a two-stage framework for multimodal personalized content generation with modal-specific bipartite interaction graphs, as depicted in Figure~\ref{fig:framework}: (1) First, we adopt a graph neural network to learn users’ modal-specific collaborative preferences, then quantize these into discrete tokens. This process is guided by designed cross-modal and preference-oriented tasks. (2) Second, we integrate the modal-specific preference tokens into downstream text and image generators, respectively, with a cross-modal consistency and personalization-aware reward.
Below, we first present modal-specific preference token learning in Section~\ref{sec:token_learning}. Then, we detail token insertion into downstream models for personalized preference integration in Section~\ref{sec:personalized_generation}. Finally, we present the cross-modal personalized consistency reward in Section~\ref{sec:reward_system}.

\subsection{Modal-specific Preference Token Learning}\label{sec:token_learning}
In this section, we focus on learning modal-specific preference tokens. First, we employ a modal-specific graph neural network to integrate modal information and collaborative filtering signals for user preference learning, which are then quantized into discrete tokens via vector quantization. To ensure tokens capture user modal preferences, we design cross-modal and preference-oriented tasks.

\subsubsection{Node Multimodal Representation Preprocessing}
Before graph aggregation, item node features in modal-specific bipartite graphs $\mathcal{G}^t$ and $\mathcal{G}^v$ are preprocessed respectively to align with downstream generators: (1) Textual modality: item text is processed via the downstream text generator, with embeddings weighted-averaged using last hidden state attention weights to form initial text features $\mathbf{h}_i^{(0),t}$. (2) Visual modality: item images are encoded using the same CLIP model~\cite{radford2021learning} as the diffusion model, yielding initial image features $\mathbf{h}_i^{(0),v}$.
This preprocessing step summarizes features, and preserves item styles. Each user node is initialized by sampling from a normal distribution parameterized by the mean and standard deviation of all modality-specific item features.

\subsubsection{Collaborative Filtering-based Encoder}
After preprocessing node features, we use graph encoders to extract modality-specific representations from graphs $\mathcal{G}^t$ and $\mathcal{G}^v$  while incorporating collaborative filtering signals~\cite{su2009survey}. 
Specifically, starting from initialized node embeddings $\mathbf{h}_u^{(0),m}$ and $\mathbf{h}_i^{(0),m}$, we apply LightGCN~\cite{he2020lightgcn}, a simple yet effective GNN for preference modeling, to perform neighbor aggregation on each $\mathcal{G}^m$ ($m\in \{t,v\}$).
At each layer, we employ LightGCN’s normalized neighborhood aggregation. After $J$ propagation layers, mean-pooling across all layer-wise embeddings yields the final aggregated user and item representations. This process can be formulated as:

\begin{equation}
    \begin{aligned}
        \mathbf{h}_{u}^{m},\mathbf{h}_{i}^{m} = \mathrm{LightGCN}(\mathbf{h}_u^{(0),m},\mathbf{h}_i^{(0),m}; \mathcal{G}^m,J), \forall m \in \{v,t\}.
    \end{aligned}
\end{equation}

\subsubsection{Modal-specific Collaborative Token Learning}
To resolve the continuous-discrete mismatch, we propose to leverage RQ-VAE~\cite{lee2022autoregressive} to generate both semantic preference tokens for users and semantic tokens for items. As a multi-stage vector quantizer, RQ-VAE produces tuples of codewords by quantizing residuals through hierarchical clustering, enabling the mapping of all user and item representations to fixed-size codebooks.
In our work, separate RQ-VAEs are constructed for the text and image modalities. For simplicity, we use $z_u$ and $z_i$ to uniformly denote the user and item representations $\mathbf{h}_{u}^{m}$ and $\mathbf{h}_{i}^{m}$ learned by graph encoders across different modalities.

Specifically, we employ a modality-specific residual quantizer consisting of $L$ levels, each with a codebook $\mathcal{B}^l:=\{e_k^l\}_{k=1}^K$, where ${e_k^l}\in \mathbb{R}^{d}$ denotes the $k$-th code vector at level $l$ and $K$ represents the size of each codebook. At the initial level, the residual is initialized as $r_{u}^1=z_u$. Then, we recursively quantize $r_{u}^l$ as follows: 
\begin{equation}
    \begin{aligned}
    c_u^l &= \mathrm{argmin}_k||r_u^l-e_{k}^l||^2, \\
r_u^{l+1}& =  r_u^l-\mathrm{Lookup}(c_u^l,\mathcal{B}^l),
    \end{aligned}
\end{equation}
where $c_u^l$ is the index of the closest code vector in $\mathcal{B}^l$, $\mathrm{Lookup}(c_u^l,\mathcal{B}^l)$ represents looking up the $c_u^l$-th code vector in $\mathcal{B}^l$.
This process thus generates a sequence of codewords:
\begin{equation}\label{eq:rqvae}
    \begin{aligned}
        (c_{u}^1,c_{u}^2,\cdots,c_{u}^L) = \mathrm{RQVAE}(z_u).
    \end{aligned}
\end{equation}

Thereby, for both text and image modalities, we obtain  $\hat{z}^t_u, \hat{z}^t_i$ and $\hat{z}^v_u, \hat{z}^v_i$. Users’ interactions are driven by multimodal preferences. Therefore, to ensure the learned tokens better capture users’ preferences at the edge level, we fuse text and image token-based representations and design a token-based preference-oriented loss function.
Specifically, we adopt the Bayesian Personalized Ranking (BPR)~\cite{rendle2009bpr} loss to enforce that observed interactions receive higher scores than unobserved ones (edge-level), formulated as follows:
\begin{equation}\label{eq:bpr_loss}
    \begin{aligned}
    \hat{z}_u& = \hat{z}_u^t \oplus \hat{z_u}^v, \quad\quad \hat{z}_i = \hat{z}_i^t \oplus \hat{z}_i^v, \\
    \mathcal{L}_{BPR} &= \sum\limits_{(u,i,i')\in \mathcal{R}} -\ln \sigma(\hat{z}_u\hat{z}_i^\top -\hat{z}_u\hat{z}_i'^\top ),
    \end{aligned}
\end{equation}
where $\oplus$ denotes concatenation; $\mathcal{R}=\{(u,i,i')\mid(u,i)\in \mathcal{R}^+,(u,i')\in \mathcal{R}^-\}$ denotes pairwise training data, $\mathcal{R}^+$ is the set of observed positive interaction edges, $\mathcal{R}^-$ is the randomly sampled negative set; and $\sigma(\cdot)$ is the sigmoid function.
Additionally, to capture node-level cross-modal alignment, we propose a cross-modal contrastive learning loss as follows:
\begin{equation}\label{eq:inter-level-contrastive}
    \begin{aligned}
    \mathcal{L}_{cm} = -ln(\sigma(\mathrm{MLP}(\hat{z}^t_i \oplus \hat{z}_i^v)-\mathrm{MLP}(\hat{z}^t_i\oplus\hat{z}_{i'}^v))).\\ 
    \end{aligned}
\end{equation}

Overall, the loss function for modal-specific preference token learning can be formulated as follows:
\begin{equation}~\label{eq:stage1_loss}
    \begin{aligned}
        \mathcal{L}_{mm-rqvae} &= \sum_{m \in \{t,v\}} \sum_{l=1}^{L} (||sg(r_{i,l}^m) - c_{i,l}^m||^2+\alpha||r_{i,l}^m - sg(c_{i,l}^m)||^2) , \\
        \mathcal{L} &= \mathcal{L}_{BPR} +  \mathcal{L}_{mm-rqvae}+\beta \mathcal{L}_{cm},
    \end{aligned}
\end{equation}
where $\beta$ denotes the hyperparameter. Upon convergence, we obtain the discrete modality-specific preference tokens for $u$ and $i$ as $P_u^m = (c_u^{m,1},\cdots,c_u^{m,L})$ and $P_i^m = (c_i^{m,1},\cdots,c_i^{m,L})$, as defined in Eq.\ref{eq:rqvae}. The learned discrete tokens naturally inherit and explainably reflect~\cite{ji2025comprehensive} the collaborative filtering effect.

\subsection{Personalized Multimodal Generation}\label{sec:personalized_generation}
To enable multimodal generation, we feed modal-specific discrete tokens $P_u^m,P_i^m$ into an LLM and a diffusion model, respectively.

\noindent\textbf{Personalized Image Generation}.
Diffusion models~\cite{rombach2022high} consist of a text encoder $c_\theta$ (e.g., CLIP~\cite{radford2021learning}) and a U-Net denoiser~\cite{ronneberger2015u}. For the text encoder, each word or sub-word is tokenized into an index that maps to a retrievable embedding vector in a predefined dictionary. During the above quantization, we obtain  user visual-level preference tokens $P_u^v = (c_u^{v,1},\cdots,c_u^{v,L})$ and item visual semantic representation tokens $P_i^v=(c_i^{v,1},\cdots,c_i^{v,L})$. We then introduce learnable placeholder tokens $U_*$ and $I_*$ in the text prompt to serve as anchors for $P_u^v$ and $P_i^v$, respectively.  During the embedding stage, we replace the embeddings of $U_*$ and $I_*$ with the modality-specific code vectors retrieved from the visual codebook $\mathcal{B}_v$ (denoted as $G^v(\cdot)$, where it represents the codebook $\text{Lookup}$ function).  A concrete prompt example is:  
$y=$ "a personalized movie poster for $\{U_*\}_{l=1}^L$, this movie $\{I_*\}_{l=1}^L$ named Titanic".  After replacing the placeholders with $G^v(P_u^v)$ and $G^v(P_i^v)$, we feed the prompt $y$ into the text encoder $c_\theta$ to obtain the contextual embedding $c_\theta(y)$. This embedding explicitly incorporates the user preference tokens $[G^{v}(c_u^{v,1}),\cdots,G^{v}(c_u^{v,L})]$ and item semantic tokens $[G^{v}(c_i^{v,1}),\cdots, G^{v}(c_i^{v,L})]$.  The U-Net then acts as a conditional denoiser for image generation. At the $t$-th denoising step, given the noisy latent $x_t$, the model predicts the noise as:
\begin{equation}
    \begin{aligned}
        \epsilon_\theta&(x_t,t,c_\theta(y)).\\
    \end{aligned}
\end{equation}

We fix both $c_\theta$ and $\epsilon_\theta$ following Textual Inversion~\cite{galimage}. After multiple denoising steps, we can obtain the generated image as $V_g$.

\noindent\textbf{Personalized Text Generation}
Similarly, we obtain user textual preference tokens $P_u^{t}$ and target item tokens  $P_i^{t}$. Besides, we design a system prompt $P_s$ to guide the text generation process.  We apply a lookup function $G^t$ (analogous to $G^v$ in the image generation) to map $P_u^t,P_i^t$ via the textual codebook, denoted as $P_u^{t}|G^t,P_i^{t}|G^t$ respectively. For example, $P_u^{t}|G^t=[G^{t}(c_u^{t,1}),\cdots,G^{t}(c_u^{t,L})]$.
After the LLM embeds  $P_s$, we sequentially concatenate $P_s,P_u^{t}|G^t,P_i^{t}|G^t$ and feed it into the LLM to generate the output text:
\begin{equation}\label{eq:final_prompt}
    \begin{aligned}
        T_g =  \mathrm{LLM}(P_s,P_u^{t}|G^t,P_i^{t}|G^t).
    \end{aligned}
\end{equation}

\subsection{Multimodal Consistent and Personalized Reward}\label{sec:reward_system}

Personalized multi-content generation is more challenging than traditional tasks, demanding explicit cross-modal consistency and preserved personalization. We thus propose a cross-modal consistent personalized reward module. For cross-modal consistency, we compute the similarity of each generated image-text pair $(V_{g},T_{g})$ using CLIP (a pre-trained multimodal network) as follows:
\begin{equation}
    \begin{aligned}
        R_\text{crs} = \mathbf{\mathrm{CLIPSim}}(V_{g},T_{g}).
    \end{aligned}
\end{equation}

To preserve personalization, we introduce a personalized reward. Taking the image modality as an example, we follow~\cite{shen2024pmg, xu2025personalized} to compute the CLIP similarity between the generated image and the user’s historically interacted images $[V_1,V_2,\cdots,V_N]$:
\begin{equation}
    \begin{aligned}
        R_{p}^v = \frac{1}{N}\sum_{i=1}^N \mathbf{\mathrm{CLIPSim}}(V_{g},V_i).
    \end{aligned}
\end{equation}

Similarly, we derive the text-based personalized reward $R_{p}^t$. Based on the training scheme of the corresponding generators, the reward-based loss can be formulated as follows:
\begin{equation}~\label{eq:image_reward}
    \begin{aligned}
    \mathcal{L}_v = ( R_{p}^v+\gamma R_\text{crs})*\mathbb{E}[\Vert \epsilon-\epsilon_\theta(x_t,t,c_\theta(y))\Vert^2],
    \end{aligned}
\end{equation}
where $x_t$ denotes the noisy version of the generated image $V_g$, $y$ is processed as illustrated in Section~\ref{sec:personalized_generation} and  $\gamma$ is a hyper-parameter.
\begin{equation}~\label{eq:text_reward}
    \begin{aligned}
        \mathcal{L}_t = -( R_{p}^t+ \gamma R_\text{crs})* \sum_{j=1}^{|y|} \log (P_{\Theta}(y_j \mid x, y_{<j})),\quad \text{where}~y=T_g,
    \end{aligned}
\end{equation}
where $\Theta$ denotes the parameters of the LLM. The total loss is then formulated as:
\begin{equation}
    \begin{aligned}
        \mathcal{L} = \mathcal{L}_t + \mathcal{L}_v.
    \end{aligned}
\end{equation}
In our work, to avoid backward gradient propagation through generators (which significantly increases training difficulty), we fix the generators’ parameters and focus on training a limited set of learned codebooks as additional generation conditions. 
This design offers not only \textit{\textbf{efficiency}} advantages but also \textit{\textbf{flexibility}}, as it enables easy extension of our approach to other pretrained models.

\section{Experiment}
In this section, we present empirical results to address the following research questions:
\textbf{RQ1} How does DPPMG perform compared with state-of-the-art personalized generation models?
\textbf{RQ2} How does DPPMG perform against baseline methods in human evaluation?
\textbf{RQ3} What are the effects of the two-stage learning and the multimodal generation module in DPPMG?
\textbf{RQ4} Why does DPPMG perform better?
\textbf{RQ5} How do hyperparameters impact the personalized generation performance?
\subsection{Experimental Settings}

\subsubsection{Dataset Description}
We conduct experiments on two datasets, focusing on movie and advertisement scenarios:
(1) \textbf{MovieLens}\footnote{https://grouplens.org/datasets/movielens}: A standard movie recommendation benchmark with user ratings on movies, where each movie includes multimodal data (textual introductions and visual posters). Following~\cite{xu2025personalized}, we adopt its latest version and retain interactions with ratings $\geq 4$.
(2) \textbf{Advertisement}: An industrial dataset containing user click data on advertisements, with each featuring multimodal data (textual copies and visual images). Detailed statistics of these two datasets are presented in Table~\ref{tab:dataset}. For both datasets, we split the samples into training, validation, and testing sets with an 8:1:1 ratio based on timestamps.

\begin{table}
    \centering
    \caption{Statistics of the datasets. ``Avg.t” denotes the average number of words for each item's textual content.}
    \begin{tabular}{c|ccccc}
    \toprule
    Dataset  & \textbf{\#user} & \textbf{\#item} & \textbf{\#interaction}  &\textbf{Avg.t}  \\
    \hline
    \textbf{MovieLens}     &598  &6,041 &45,276 &25.58\\
    \textbf{Advertisement} & 1,921& 4,229 &50,010 &9.21\\
    \bottomrule
    \end{tabular}
    
    \label{tab:dataset}
\end{table}
\subsubsection{Implementation Details}\label{apx:implement_detail}

All experiments were conducted on a single NVIDIA A100 GPU.
We adopt LLaMA3-8B-Instruct~\cite{grattafiori2024llama} as the LLM backbone and Stable Diffusion v1.5~\cite{rombach2022high} as the image generator for all baselines, with the sole exception of Pigeon, whose default configuration employs Stable Diffusion XL~\cite{podellsdxl}.  For our framework, we default the layers of LightGCN to 2, as it is not the focus of our work.  The residual quantization codebook comprises $L = 4$ layers and codebook size $K=96$ for each layer. Sec.\ref{sec:hyperparameter} further investigates their impact.  The loss terms $\beta$ and $\gamma$ in Eq.~\ref{eq:stage1_loss} are set to 0.1 and 0.5, respectively.
For the first-stage training, we set the batch size as 1024 and learning rate as 1e-3; For the second-stage training, we set the batch size as 8 and the learning rate as 1e-5. Both stages are optimized using the Adam optimizer~\cite{kingma2014adam}. For inference, each sample takes about \textbf{4 seconds} to generate an image. 
As shown in Table~\ref{tab:param_an}, in contrast to the large total parameter counts of LLM/diffusion models, DPPMG has only a small fraction of trainable parameters, leading to relatively low computational overhead. Specifically, the image and text codebooks contain merely $4\ \text{quantizers} \times 96\ \text{codes} \times 768\ \text{dimensions} = 294,912$ and $4 \times 96 \times 4096 = 1,572,864$ parameters, respectively.
\begin{table}[]\small
    \centering
    \caption{Model parameters and trainable ratio of DPPMG}
    \label{tab:param_an}
    \begin{tabular}{c|ccc}
        \toprule
          &\textbf{Trainable Parameters}&\textbf{DM/LLM Parameters}&\textbf{Ratio} \\
         \hline
        \textbf{Image}&1,640,448& 1,066,236,843&0.154\%\\
        \textbf{Text} &36,012,032 &8,030,277,632&0.448\% \\
         \bottomrule
    \end{tabular}
\end{table}

\subsubsection{Evaluation Metrics}

For DPPMG’s multimodal generation evaluation, we adopt the following metrics:
\textit{(1) Personalized Image Generation}.
Following prior work~\cite{shen2024pmg,shilova2023adbooster,xu2025personalized}, we evaluate personalization and semantic alignment with reference images via three categories: (a) \textbf{Semantic Metrics}: CLIP/DINO Image Scores (CIS/DIS), defined as the cosine similarity of visual features from CLIP~\cite{radford2021learning}/DINOv2~\cite{oquab2023dinov2}; and CLIP Score (CS), measuring cross-modal alignment between generated images and reference text; (b) \textbf{Perceptual Metrics}: LPIPS~\cite{zhang2018unreasonable} and MS-SSIM~\cite{wang2003multiscale} for fine-grained visual similarity; (c) \textbf{Fidelity}: Fréchet Inception Distance (FID)~\cite{heusel2017gans} for generated image distribution consistency. \textbf{Overall} denotes the F1-score combining historical CIS and reference CS for overall performance.
\textit{(2) Personalized Text Generation}. We focus on personalization and semantic alignment with reference text (movie introduction/advertisement copy) via semantic similarity: (a) \textbf{Precision}: BLEU~\cite{papineni2002bleu} relies on n-gram matching, with BLEU-1 for word accuracy and BLEU-4 for fluency); (b) \textbf{Recall}: ROUGE~\cite{lin2004rouge} emphasizes content coverage, with ROUGE-N for n-gram overlap and ROUGE-L for longest common subsequence coherence.
\textit{(3) Multimodal Content Generation}.
We compute the \textbf{CLIP Consistency Score} (CCS) as the cosine similarity of CLIP-extracted features from generated images and texts, to evaluate cross-modal consistency.

\subsection{Overall Performance Comparison (RQ1)}\label{sec:overall_performance}
Given the absence of personalized multimodal generation models, we evaluate the performance of DPPMG on both personalized image generation and text generation tasks to examine its effectiveness.
\subsubsection{Personalized Image Generation Task}
We compare DPPMG with the following personalized image generation baselines:
\begin{itemize}[leftmargin=*,itemsep=2pt,topsep=0pt,parsep=0pt]
\item \textbf{Textual Inversion (TI)}~\cite{galimage}: Employs word embeddings to learn user preferences, which are combined with textual instructions to guide diffusion-based text-to-image generation.
\item \textbf{PMG}~\cite{shen2024pmg}: Converts user-interacted/reference images into text descriptions, extracts keyword-based preferences via pre-trained LLMs, and conditions image generators on the combination of preference keywords and implicit embeddings.
\item \textbf{Qwen-VL}~\cite{Qwen2VL}: An MLLM for image feature extraction and visual reasoning, used to extract user preferences via keywords from interacted images and enable image generation with external text-to-image generators.
\item \textbf{LaVIT}~\cite{jin2023unified}: An MLLM that converts images into discrete tokens for reasoning and generates visual tokens for image generation.
\item \textbf{I-AM-G}~\cite{wang2024g}: Obtains user preference textual keywords from multimodal sparse interactions using an MLLM, and integrates these keywords with relevant content for generation.
\item \textbf{Pigeon}~\cite{xu2025personalized}: Leverages LaVIT’s visual understanding and reasoning capabilities to extract user visual preferences, with a token-level masked generation module for noisy historical images.
\end{itemize} 
Table~\ref{tab:image_performance} presents results on both datasets, where we observe that:
\begin{table*}[]\small
\caption{
    Overall performance of personalized image generation task in both scenarios. Baselines labeled with "*" are pre-trained models.  The best results are highlighted in bold, while the second-best results are underlined.}\label{tab:image_performance} 
    \centering
    \begin{tabular}{c|c|c|ccccc|ccc|c}
    \toprule
            \multirow{2}{*}{Dataset}&\multirow{2}{*}{Methods}&\multirow{2}{*}{Overall}&\multicolumn{5}{c|}{Personalization} &\multicolumn{3}{c|}{Semantic Alignment} & Fidelity  \\
           \cline{4-12}
          &&&CS$\uparrow$ &CIS$\uparrow$ & DIS$\uparrow$ &LPIPS$\downarrow$ &MS-SSIM$\uparrow$ & CS $\uparrow$ &CIS$\uparrow$ & DIS$\uparrow$ &FID$\downarrow$ \\
          \cline{1-12}
          \multirow{8}{*}{\#MovieLens}& TI &28.05 & 12.18 &33.98&\underline{22.03}&0.8022&0.0574&22.12&\underline{51.74}&36.13&96.74 \\

           & PMG&25.47& 10.17&28.61&16.38&0.7889& 0.0598&22.33&43.19&30.24&99.98\\

          &Qwen-VL* &27.66&11.05&32.67&18.88& 0.8001&0.0603&\underline{22.65} &\textbf{52.14}&39.69&  84.48\\ 
          
          &LaVIT* &24.51&11.52&27.57&21.43&0.7935&0.0670&21.45 &38.96&33.11& 66.22\\
          &I-AM-G &27.58&\underline{13.98}&32.95&20.44 &0.7653 &0.0632 & 22.21 &47.91&40.98&67.20 \\
          &Pigeon &\underline{28.16}&13.94&\underline{34.21}&19.37&\underline{0.7631}&\textbf{0.0826}&22.10 &49.71&\underline{41.19}& \underline{59.51}\\
          \cline{2-12}
           
         &DPPMG& \textbf{29.49}&\textbf{14.90}&\textbf{35.64}&\textbf{23.07}&\textbf{0.7628}&\underline{0.0782}&\textbf{23.33}&50.39&\textbf{42.90}&\textbf{57.38}\\
        \hline
        \hline
        \multirow{8}{*}{\#Advertisement}& TI &26.07& 20.73&30.69&12.85&0.7907&0.1406&21.54&44.12&\textbf{32.71}&94.19 \\

          & PMG& 26.33&20.69&31.18&12.22&0.7880&0.1376&21.48&42.50&30.64&98.43\\

          &
          Qwen-VL* &27.34 & 21.24 &32.61&12.01& 0.7836&0.1439&22.07&41.68&30.75&94.85\\  
          &LaVIT* &25.66&22.65 &32.10&13.71&\textbf{0.7776}&0.1448&19.21&40.47 & 27.50&97.65\\
          &I-AM-G &28.85&22.46&35.64&13.30&0.7853&\underline{0.1481}&22.05&\underline{46.45}&\underline{31.48}&90.77 \\

          &Pigeon &\underline{29.44}&\underline{22.67}&\underline{36.55}&\underline{14.22}&\underline{0.7821}&\textbf{0.1510}&\underline{22.32}&45.76&30.41&\underline{83.42}\\
          \cline{2-12}&DPPMG&\textbf{30.48}&\textbf{23.08}&\textbf{37.83}&\textbf{14.68}&0.7846&0.1461&\textbf{23.12}&\textbf{47.29}&27.90&\textbf{79.83}\\

        \bottomrule
        
    \end{tabular}
    
\end{table*}

\begin{itemize}[leftmargin=*,itemsep=2pt,topsep=0pt,parsep=0pt]
   \item DM-based TI can have comparable semantic alignment by directly using the reference image’s textual description for text-to-image generation. However, not all visual features in the interaction history reflect user preferences, thus introducing noisy signals that hinder personalized image generation.
   \item PMG uses LLM to infer user preferences in textual form  to guide image generation. However, user visual preferences may be lost when converted to text, thus limiting personalization.
   \item Qwen-VL, LaVIT, I-AM-G and Pigeon take images as input but rely on generative models not specialized for user preference capture, leading to suboptimal personalized image generation.

   \item  DPPMG outperforms baselines on most personalization and fidelity metrics across both scenarios (such as Personalization-CIS peaking at 35.64) while maintaining comparable semantic alignment, demonstrating that its discrete tokens can effectively guide diffusion models to produce personalized images.
\end{itemize}

\subsubsection{Personalized Text Generation Task} 
\begin{figure}
    \centering 
    \subfigure[Preference token learning applied in personalized text generation methods]
    {
    \includegraphics[width=0.45\linewidth]{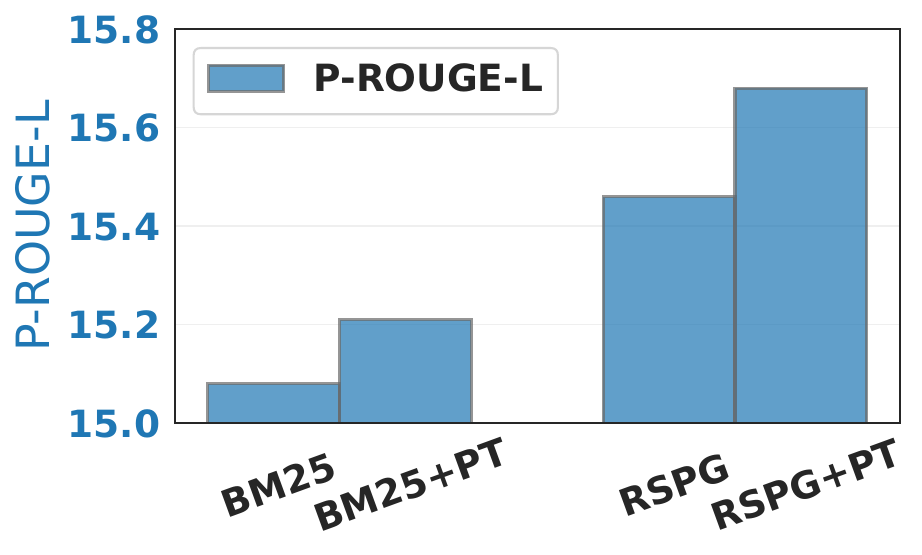}\label{fig:text_per}
    }
    \subfigure[Preference token learning applied in different backbones]
    {
    \includegraphics[width=0.45\linewidth]{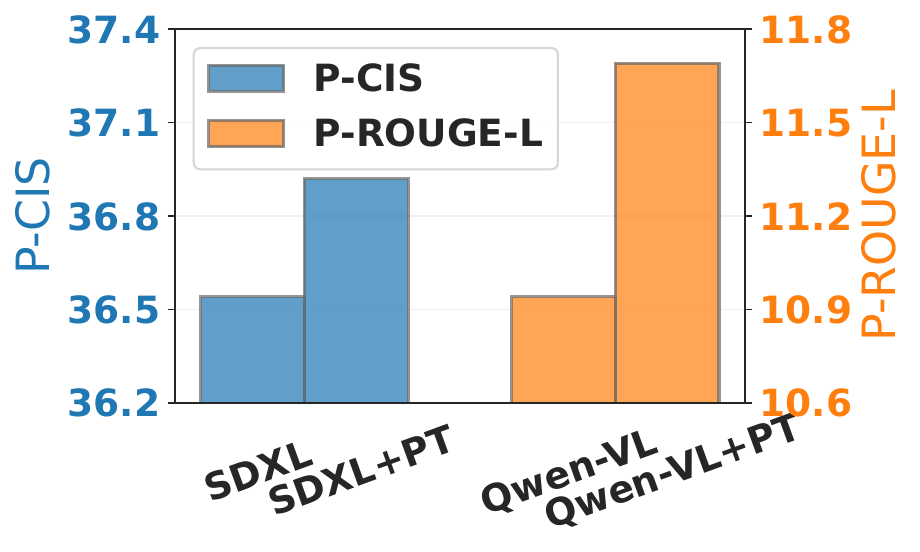}\label{fig:plug_in1}
    }
    \caption{Plug-in analysis of preference token learning.
    }\label{fig:plug_all}

\end{figure}

We compared DPPMG with LLMs (\textbf{LLaMA*}~\cite{grattafiori2024llama}, our backbone without learned tokens) and MLLMs (\textbf{Qwen-VL*}~\cite{Qwen2VL}, the same prompts plus reference images for dense visual conditioning).
Furthermore, we fine-tune them on the text completion task (causal language modeling) via LoRA~\cite{hu2021lora}. 
From the results in Table \ref{tab:text_results}, we can observe:
\begin{itemize}[leftmargin=*,itemsep=2pt,topsep=0pt,parsep=0pt]
    \item   DPPMG outperforms baselines on most metrics, especially personalized ones, and this advantage persists over the backbone LLaMA* (without our learned tokens), indicating that our learned tokens enhance personalized text generation.
    \item Fine-tuning improves the text semantic alignment of Qwen-VL/LLaMA over pre-trained versions. However, it yields no personalized performance gains without preference modeling.
   
\end{itemize}
\begin{table*}
\small
  \caption{Overall performance of personalized text generation task in both datasets.} \label{tab:text_results}
  \begin{tabular}{c|c|ccccc|ccccc}
    \toprule
     \multirow{2}{*}{Dataset}&\multirow{2}{*}{Model}&\multicolumn{5}{c|}{Personalization} &\multicolumn{5}{c}{Semantic Alignment} \\
     \cline{3-12}
    && ROUGE-1 & ROUGE-2 & ROUGE-L& BLEU-1&BLEU-4 &  ROUGE-1 & ROUGE-2 & ROUGE-L& BLEU-1&BLEU-4  \\
    \cline{1-12}
    \multirow{5}{*}{\#MovieLens} &LLaMA*&14.7955&0.7265&\underline{10.9469} &\underline{9.8443}&0.1542&13.3755&0.5543&10.2332&8.7781  &0.1302  \\
    &LLaMA & 14.0268 &0.8291&10.5956&9.2007&0.1835&
    14.1676 &0.5676 &10.5234 &\underline{9.5443}
    &0.1471 \\
    &Qwen-VL* & \underline{14.8117}& 0.7829 &10.9407&7.1190&0.1703 &14.5568&0.7047&10.8060&7.1190&\underline{0.1703}\\
    &Qwen-VL &14.5322&\underline{0.8309}&10.4523&8.2234&\underline{0.1960}&\underline{15.1335}&\textbf{0.7647} &\underline{11.0952}&8.4671&\textbf{0.1835}\\
    \cline{2-12}
    &DPPMG& \textbf{16.0997} &\textbf{0.9058}&\textbf{11.8643}&\textbf{10.9744}&\textbf{0.2061}&\textbf{15.4185}&\underline{0.7182}&\textbf{11.4033}&\textbf{10.5584}&0.1687\\
    \hline
    \hline
    \multirow{5}{*}{\#Advertisement}
     &LLaMA*&15.1182 &\underline{7.3822}&\underline{11.9924} &1.5194&0.5343&8.5882&2.8002&7.3677&0.8301&0.1632  \\
   & LLaMA &\underline{15.1275} &7.3513&11.1786 &1.4075&\underline{0.6240}  &8.9134 &\underline{2.8446} &\underline{7.6440} &1.3162 &\underline{0.3536  }\\
    &Qwen-VL*&12.6661 & 4.5622&10.6788&\textbf{2.9944} &0.2386&8.8257&2.1175&7.4759&1.5853&0.1538 \\
    &Qwen-VL &13.5630 & 4.7021 &10.9505&2.3082&0.2782 & \textbf{9.3221}&2.7092 & 7.5048&\textbf{2.0367}&0.2487 \\
    \cline{2-12}
    &DPPMG& \textbf{15.6615}&\textbf{ 7.6645}&\textbf{13.7366}& \underline{2.7576}&\textbf{0.9913}&\underline{9.0362}&\textbf{3.0180}&\textbf{7.9142}&\underline{1.7219}&\textbf{0.3726}\\
    \bottomrule
\end{tabular}
\end{table*}
Note that existing personalized text generation works~\cite{salemi2023lamp,salemi2024optimization,tan2024personalized}, which primarily augment prompts via retrieval from user historical interactions, are orthogonal to our discrete token-insertion method. Given their distinct operational stages, we propose hybrid extensions by integrating preference-token learning (shortened as PT) into two such methods: BM25~\cite{salemi2023lamp} and RSPG~\cite{salemi2024optimization} based on our LLaMA backbone. The results in Figure~\ref{fig:text_per} confirm that \textbf{preference token learning can be seamlessly incorporated into orthogonal retrieval-augmented frameworks}, enhancing personalized text generation with collaborative filtering cues.

Besides, to further verify its plug-in effectiveness, we apply the preference-guided token learning process to additional backbones of Stable Diffusion XL~\cite{podellsdxl} (shortened as SDXL) and Qwen-VL~\cite{Qwen2VL}, with results presented in Figure~\ref{fig:plug_in1}. 
The experimental results demonstrate that integrating our method into SDXL and Qwen-VL can consistently improve personalized image and text generation performance, respectively.
They further confirm the generalizability and reliable \textbf{plug-in effectiveness} of our proposed strategy across different backbones.

\subsection{Human Evaluation (RQ2)} 
We conducted a human study comparing DPPMG with two competitive baselines: Pigeon for image and Qwen-VL for text. Thirty participants completed three ranking tasks on fifty movie cases: two tasks evaluated personalization, with participants ranking generated images and text by click likelihood based on historical items; the third focused on multimodal alignment, with participants ranking samples by cross-modal consistency between generated images and text.  For personalized performance, \textbf{62.3\%} of participants preferred DPPMG-generated images to those from Pigeon, and \textbf{64.2\%} favored its generated texts over those from Qwen-VL. For multimodal alignment, \textbf{68.1\%} found DPPMG produced more cross-modal consistent content. The results demonstrate DPPMG outperforms baselines in both personalization and multimodal alignment.

\subsection{Ablation Study (RQ3)}\label{sec:core_component}
\subsubsection{Impact of Two-Stage Learning}
We conducted the following ablation studies to validate discrete preference token learning and cross-modal consistency personalized reward:
(1) \textit{Base}: Directly generates text with the LLM and images with the diffusion model, without the learned token.
(2) \textit{Base+Preference token}: Generates text and images using the tokens learned in the first stage.
(3) \textit{Base+Preference token+Unimodal reward}: Learns tokens in the first stage and further fine-tunes them using a unimodal personalized reward.
(4) \textit{Base+Preference token+Cross-modal reward}: Similarly learns tokens but fine-tunes them with the cross-modal personalized consistency reward.
From the results in Table~\ref{tab:ablation}, we observe:
\begin{itemize}[leftmargin=*,itemsep=2pt,topsep=0pt,parsep=0pt]
    \item  Compared with \textit{Base}, \textit{Base+Preference Token} can improve personalization for both image and text generation tasks, demonstrating that the learned tokens reflect users' preferences.
    \item As expected, unimodal rewards boost personalization (personalized metrics peaking at 35.71, 23.42, 12.02), while cross-modal rewards lift the fifth metric to 22.43 for better consistency.
    \item Jointly fine-tuning tokens with both unimodal and cross-modal rewards allows our model to strike a balanced trade-off, generating multimodal content that is both personalized and consistent.
\end{itemize}
\begin{table*}[]\small
    \centering
    \caption{Ablation of preference token and cross-modal personalized reward in DPPMG. In the metrics, "P" represents personalization while "A" represents the alignment with reference text or image. CCS metric measures the cross-modal alignment.}\label{tab:ablation}
    \begin{tabular}{c|l|cc|cc|c}
    \toprule
    \multirow{2}{*}{Dataset}&\multirow{2}{*}{Ablation}&\multicolumn{2}{c|}{Image} &\multicolumn{2}{c|}{Text} & Cross-Modal \\
    \cline{3-7}
        &&P-CIS & A-CS &P-ROUGE-L&A-ROUGE-L & CCS\\
    \cline{1-7}
       \multirow{5}{*}{\#MovieLens}
       &\textit{Base} &32.07  &21.47&10.95 &10.23 &19.13 \\
        &\textit{Base+Preference token}  &33.89 &22.13 &11.41&10.40&19.15\\
        &\textit{Base+Preference token+Unimodal reward}  &\textbf{35.71} &\textbf{23.42} &\textbf{12.02}&\underline{11.34}&19.01 \\
        &\textit{Base+Preference token+Cross-modal reward}  &33.98 &21.50 &11.24&11.06&\textbf{22.43 } \\

        &Ours(final) &\underline{35.64} &\underline{23.33} &\underline{11.86}&\textbf{11.40}&\underline{21.32} \\
    \hline
    \multirow{5}{*}{\#Advertisement}&\textit{Base} &35.58 &21.46 &11.99 &7.37&20.35 \\
    &\textit{Base+Preference token}  &36.39 &21.73 &13.03&7.21&21.32\\
    &\textit{Base+Preference token+Unimodal reward}  &\textbf{37.95} &\textbf{23.33} &\textbf{14.26}&\underline{7.46}&19.81  \\
    &\textit{Base+Preference token+Cross-modal reward}  &35.98 &22.76 &12.13&6.99&\textbf{22.71}  \\
    &Ours(final) & \underline{37.83} &\underline{23.12} & \underline{13.74}&\textbf{7.91}&\underline{22.10} \\
    \bottomrule
        
    \end{tabular}
    
\end{table*}

\subsubsection{Impact of Multi-Modal Generation}
\begin{figure}
    \centering 
    \subfigure[image-only vs. multimodal]
    {
    \includegraphics[width=0.45\linewidth]{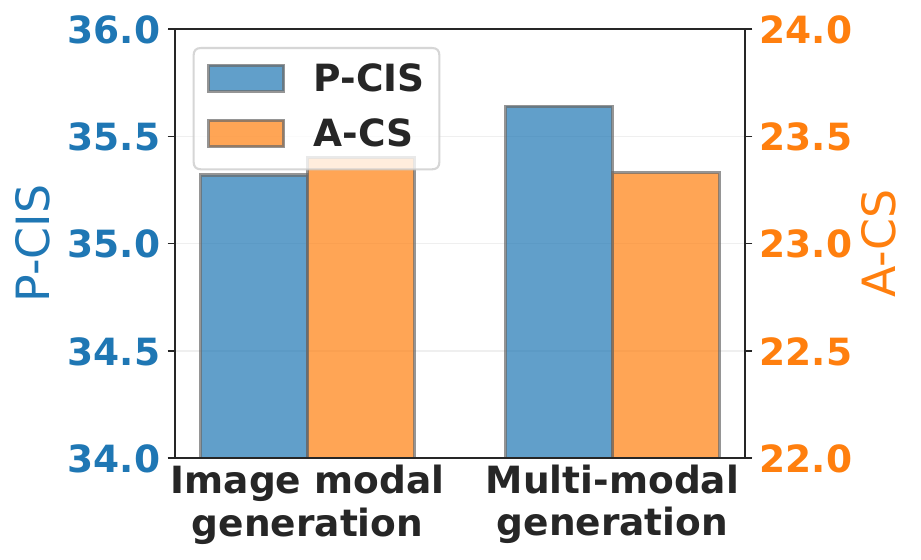}\label{fig:image_modal}
    }
    \subfigure[text-only vs. multimodal]
    {
    \includegraphics[width=0.45\linewidth]{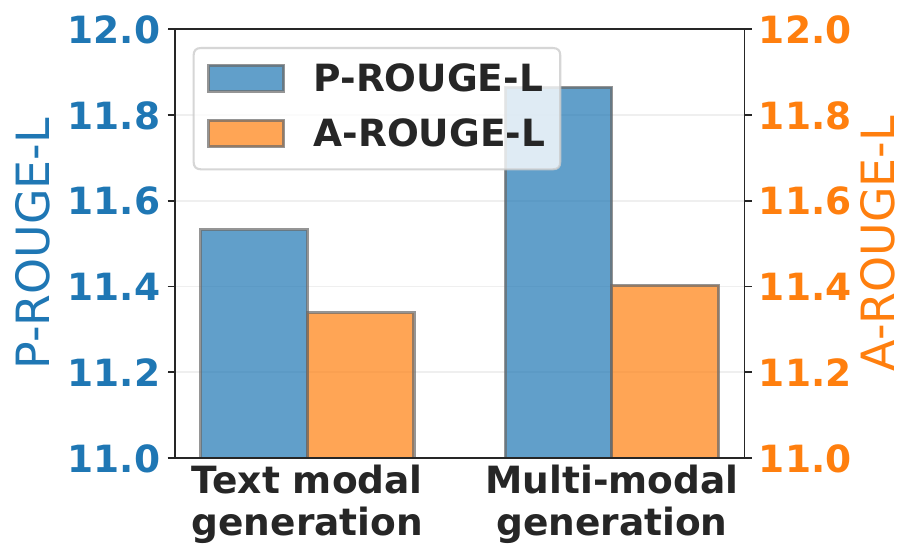}\label{fig:text_modal}
    }
    \caption{Unimodal generation vs. Multimodal generation.}\label{fig:multi-unimodal}

\end{figure}

To verify the impact of multimodal modeling, we  conduct an ablation study by decomposing DPPMG into two unimodal variants (image-only and text-only) by excluding all multimodal components. For the image-only variant, we only retain $\hat{z}^v_u$ and $\hat{z}^v_i$ to formulate the token-based preference loss in Eq.~\ref{eq:stage1_loss} during the first training stage;
In the second stage, we discard the cross-modal consistency reward term in Eq.~\ref{eq:image_reward}. The text-only variant follows the similar logic. From the ablation results in Figure~\ref{fig:multi-unimodal},  personalized performance drops for both unimodal settings. This can be attributed to the fact that user interactions are typically driven by multimodal content; with only unimodal information, the preference tokens fail to capture comprehensive user preferences, thus leading to suboptimal personalized performance.

\subsection{In-depth Analysis (RQ4)}

\begin{table}[]\small
\caption{Aggregate metrics for movie posters sharing prefix length. Subtree size counts movies sharing that prefix. }
    \centering
    \begin{tabular}{c|cr}
    \toprule
        \textbf{Share prefix length} &\textbf{Avg pairwise similarity} & \textbf{Subtree size}  \\
        \hline
        1 &0.4967&(12-129) \\
        2&0.6047&(1-12) \\
        3&0.9757&(1-3) \\
        4&0.9973 &(1-2)\\
        \bottomrule
    \end{tabular}
    
    \label{tab:itemtoken}
\end{table}

\begin{figure}
    \centering
    \includegraphics[width=0.7\linewidth]{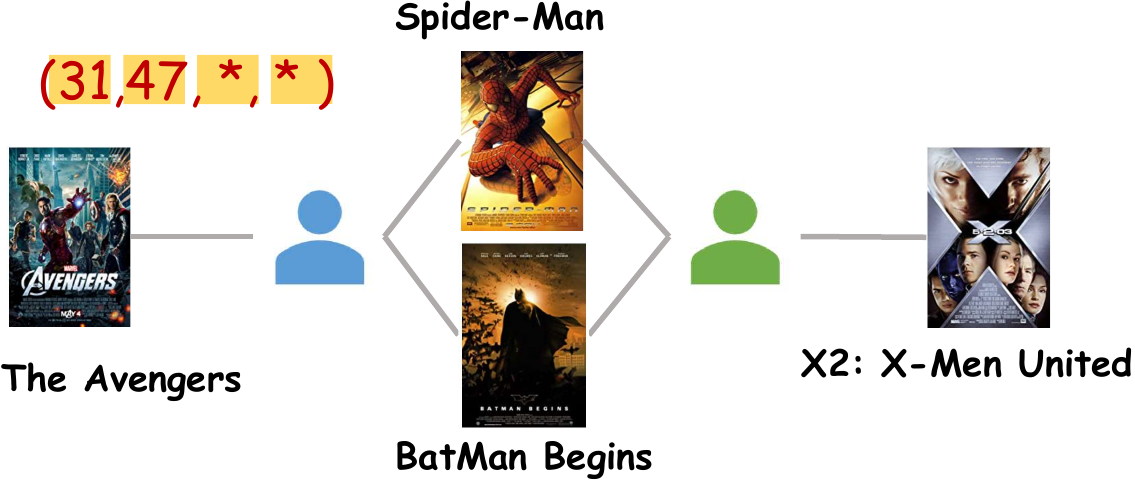}
    \caption{A case of collaborative effect learned by tokens. Shared token prefix captures user and item similarity.}
    \label{fig:collaborative}
\end{figure}

To enhance personalized generation, we integrate LightGCN and RQVAE: LightGCN learns modal-specific user/item representations from collaborative interaction data, while RQVAE discretizes these into hierarchical tokens via coarse-to-fine residual quantization. This design captures the layered structure of user preferences and item features, with higher-level tokens encoding broad attributes and lower-level ones fine-grained details, enabling nuanced personalization. We demonstrate this via image token learning in movie scenarios.
\textbf{Hierarchical Item Similarity}:
Table~\ref{tab:itemtoken} shows items sharing longer token prefixes (e.g., items with tokens $(1, 2, 3, 4)$ and $(1, 2, 6, 7)$ share a prefix of length 2) have higher pairwise cosine similarity and smaller subtree sizes. This confirms a hierarchical token structure reflecting graded item similarity.
\textbf{Collaborative Filtering Learning}: We aim to capture collaborative filtering signals for better modeling of user preferences. To illustrate this, we present an example in Figure~\ref{fig:collaborative} where users and items share the first two tokens, with analysis from two aspects: \textit{(1) User-based}: Both users interacted with \textit{Spider-Man} and \textit{Batman Begins}, indicating similar interaction patterns. Thereby, the left user may enjoy \textit{X2: X-Men United}  while the right user may favor \textit{The Avengers}. Their shared initial token prefix confirms that tokens encode user-based collaborative behavior by aligning users with similar interaction histories.
\textit{(2) Item-based}: Building on the hierarchical item similarity described above in Table~\ref{tab:itemtoken}, shared prefixes capture item affinities. This allows the model to identify items similar to user-interacted ones, enabling item-based collaborative filtering.

\subsection{Impact of Hyperparameter (RQ5)}\label{sec:hyperparameter}
In this section, we conduct experiments to investigate the impact of key hyperparameters.

\begin{figure}
    \centering 
    \subfigure[Personalized performance w.r.t the layer of codebook]
    {
    \includegraphics[width=0.45\linewidth]{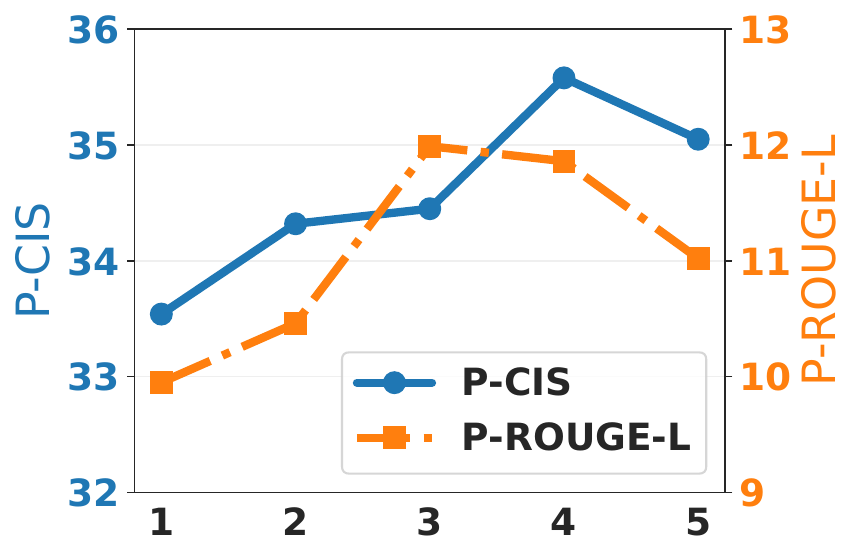}\label{fig:codelayer}
    }
    \subfigure[Personalized performance w.r.t the size of codebook]
    {
    \includegraphics[width=0.45\linewidth]{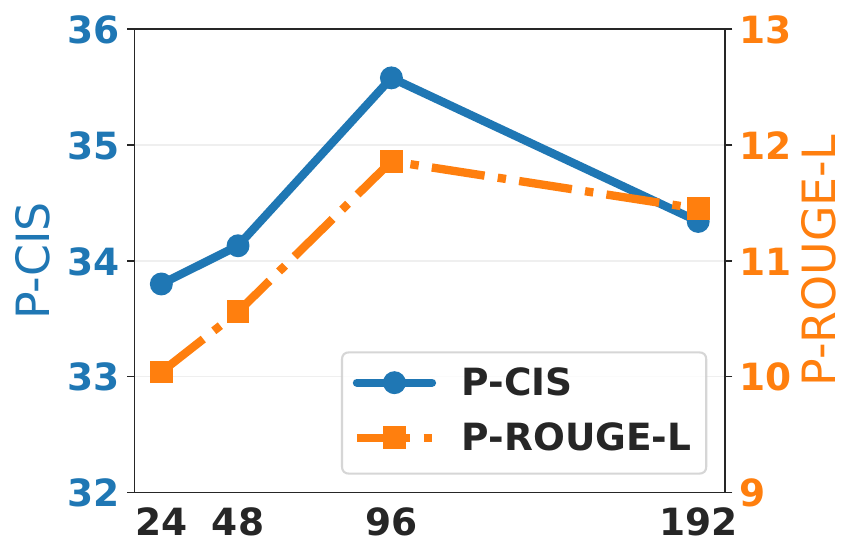}\label{fig:codesize}
    }
    \caption{Impact of hyperparameters.}\label{fig:hyparameter}

\end{figure}

\subsubsection{Impact of Codebook Layer}
We vary $L$ within $\{1,2,3,4,5\}$ with $K=96$.  Figure~\ref{fig:codelayer} shows that performance improves and then declines as $L$ increases. When $L<3$, the codebook capacity is insufficient, limiting its representational ability. Specifically, $L =3$ yields optimal performance for personalized text generation, while $L=4$ is optimal for personalized image generation. This discrepancy may arise because image features are more complex than textual features, requiring a larger $L$ to capture fine-grained nuances. As $L$ continues to increase, features may be decomposed into redundant high-order residuals, leading to overfitting to the training data and reduced generalization.

\begin{figure}
    \centering
    \includegraphics[width=0.9\linewidth]{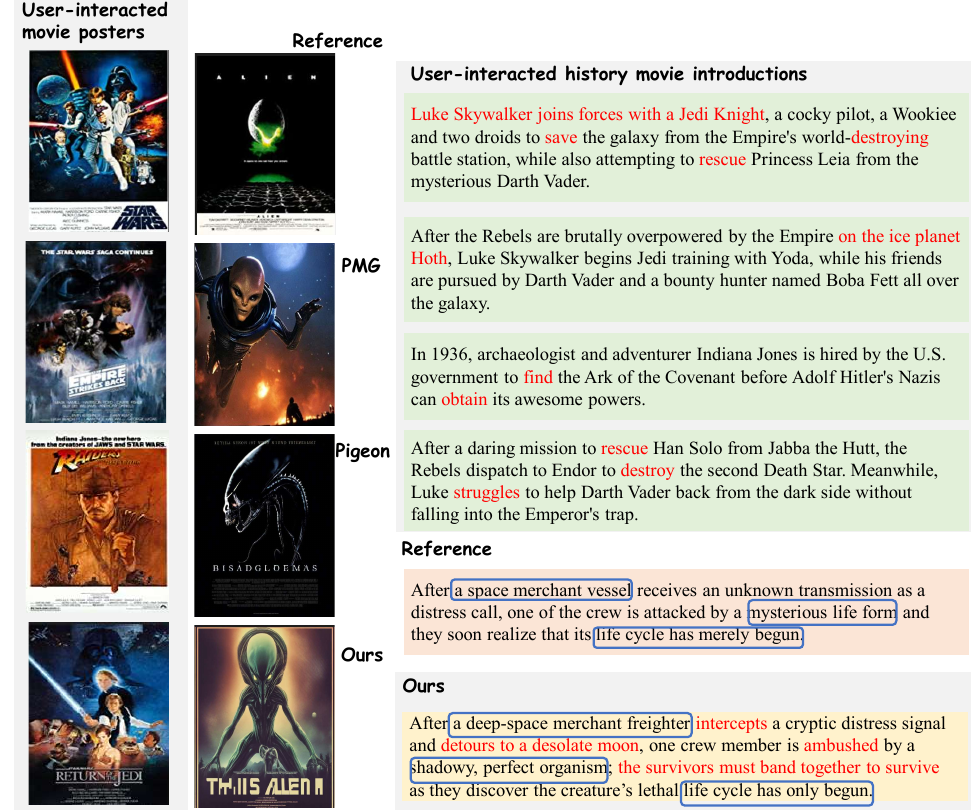}
    \caption{An example of generated image and text for movie.}
    \label{fig:case}
\end{figure}
\subsubsection{Impact of Codebook Size}
We vary $K$ in $\{24,48,96,192\}$ with fixing $L=4$. The results in Figure~\ref{fig:codesize} indicate a performance trend of first increasing and then decreasing. When $K<96$, the codebook capacity per layer may be insufficient to cover the vector space.  $K =96$ achieves optimal performance for both image and text generation tasks.  However, as $K$ continues to increase, performance declines due to training instability caused by the exponential moving average training scheme~\cite{van2017neural} in RQVAE.

\subsection{Case study}
In this section, we present an example of a DPPMG-generated image–text pair, along with four user historical interacted samples. For image generation, we compare DPPMG against two personalized baselines (PMG and Pigeon), as illustrated in Figure~\ref{fig:case}.
(1) Image modality: The DPPMG-generated poster for the movie \textit{Alien} achieves high semantic alignment with the reference poster by emphasizing the original black-and-green palette and glossy, ribbed carapace to preserve its mysterious style. Simultaneously, it aligns with the user’s preference for soft, luminous outlines and slightly cartoon-styled characters of historical interaction. (2) Text modality: The generated introduction matches the style of the user’s historical introductions—for instance, the rapid action cadence of the verb chain \textit{intercepts, detours, ambushed, band together, survive} is similar to \textit{rescue, destroy, struggle} in historical interacted texts.  
It also maintains strong semantic consistency with the reference text, preserving key elements such as \textit{one crew member is ambushed} and \textit{deep-space merchant freighter}. (3) Cross-modal consistency: The phrase a shadowy, perfect organism in the generated text corresponds to the mysterious alien silhouette in generated image.

\begin{figure}
    \centering {\includegraphics[width=0.9\linewidth]{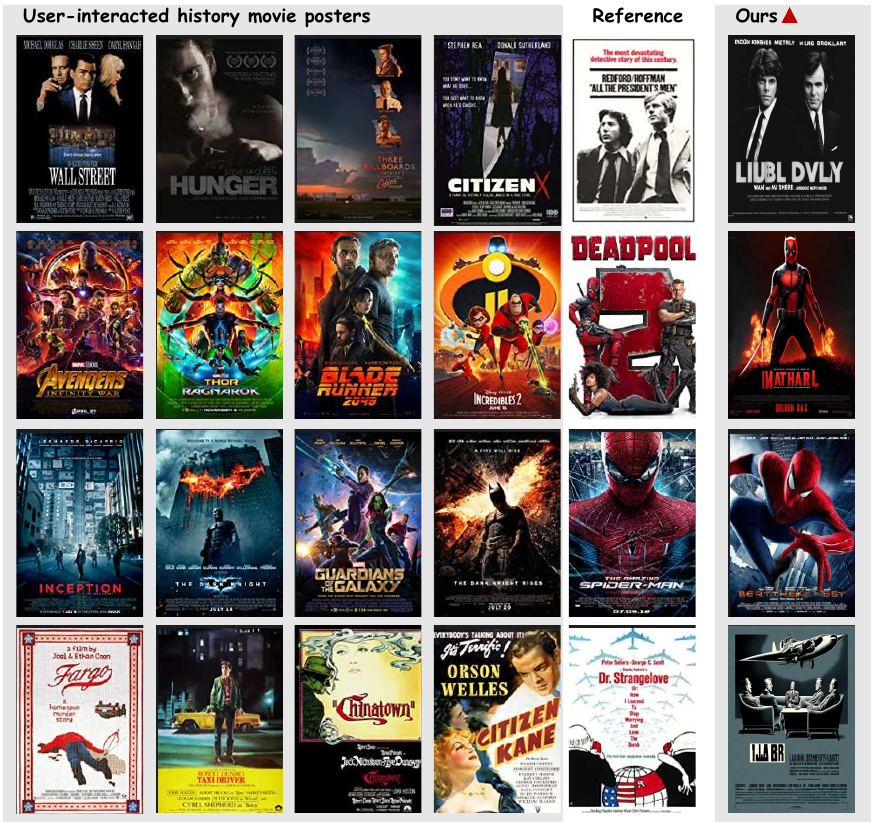}
    }{\includegraphics[width=0.9\linewidth]{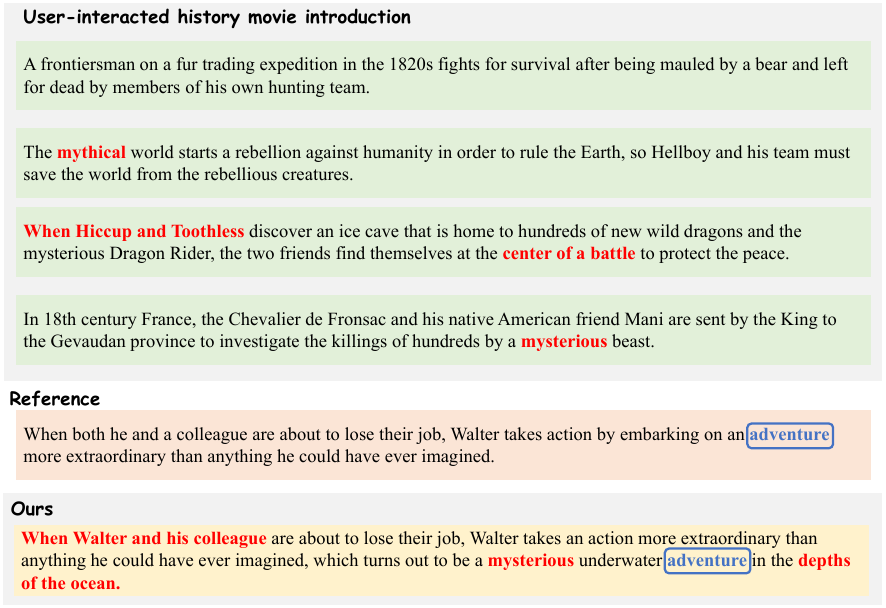}
    }
    \caption{More examples of the generated personalized movie posters and text introductions.}\label{fig:other_cases}
\end{figure}

Furthermore, we present additional examples of generated movie posters and movie introductions in Figure~\ref{fig:other_cases}, respectively. DPPMG-generated posters effectively mirror users’ preferences through character-centered designs, dynamic compositions, and color palettes that align with each user’s unique taste, while its generated introductions reflect these preferences via theme keyword focus and plot tendencies consistent with the user’s personal preferences.

\section{Conclusion}

This paper introduces a novel two-stage discrete token learning framework for personalized multimodal generation. It first discretizes users’ modal-specific preferences into discrete tokens via a graph neural network-based quantization stage. These tokens are then injected into downstream generators and refined with a personalized cross-modal consistency reward to ensure both consistency and personalization. Experimental results and human evaluation demonstrate its effectiveness in generating personalized and consistent multimodal content. Further experiments also verify its plug-in effectiveness, and show that the learned tokens can demonstrate classical collaborative filtering effects.

This work marks an initial attempt at personalized multimodal generation, paving the way for several promising directions: (1) adapting DPPMG to leverage a more advanced unified generative model, which may help guide personalized text and image generation toward greater consistency; (2) expanding the scope of personalized generation to more diverse multimodal combinations, such as incorporating audio alongside text and images.

\begin{acks}
This work is partly supported by the National Natural Science Foundation of China (No. 62306255, 92370204), the National Key Research and Development Program of China (No. 2023YFF0725001), the Guangdong Basic and Applied Basic Research Foundation (Grant No.2023B1515120057), the Key-Area Special Project of Guangdong Provincial Ordinary Universities (2024ZDZX1007).
\end{acks}
\bibliographystyle{ACM-Reference-Format}
\balance
\bibliography{ref}

\end{document}